\documentclass{article}
\usepackage{graphics}
\pdfoutput=1
\usepackage{amssymb,amsfonts,amsmath}
\usepackage[serbian,english]{babel}
\usepackage{multirow}
\usepackage{rotating}
\usepackage{fullpage}
\usepackage[utf8]{inputenc}

\newcommand\institute[1]{\newcommand\theinstitute{#1}}

\title{The thermodynamics of human reaction times}
\author{Ferm\'{\i}n MOSCOSO DEL PRADO MART\'IN}
\institute{Laboratoire de Psychologie Cognitive ({UMR}-6146), Centre National de la Recherche Scientifique \& Aix--Marseille Universit\'e I, Marseille, France}

\begin{document}
\maketitle

\begin{abstract} I present a new approach for the interpretation of reaction time (RT) data from behavioral experiments. From a physical perspective, the entropy of the RT distribution provides a model-free estimate of the amount of processing performed by the cognitive system. In this way, the focus is shifted from the conventional interpretation of individual RTs being either long or short, into their distribution being more or less complex in terms of entropy. The new approach enables the estimation of the cognitive processing load without reference to the informational content of the stimuli themselves, thus providing a more appropriate estimate of the cognitive impact of different sources of information that are carried by experimental stimuli or tasks. The paper introduces the formulation of the theory, followed by an empirical validation using a database of human RTs in lexical tasks (visual lexical decision and word naming). The results show that this new interpretation of RTs is more powerful than the traditional one. The method provides theoretical estimates of the processing loads elicited by individual stimuli. These loads sharply distinguish the responses from different tasks.  In addition, it provides upper-bound estimates for the speed at which the system processes information. Finally, I argue that the theoretical proposal, and the associated empirical evidence, provide strong arguments for an adaptive system that systematically adjusts its operational processing speed to the particular demands of each stimulus. This finding is in contradiction with Hick's law, which posits a relatively constant processing speed within an experimental context.\\
{\bf keywords:} cognition $|$ entropy $|$ lexical decision $|$ reaction time $|$ word naming
\end{abstract}

\hyphenation{neg-en-tro-py}


Ever since its introduction by Donders \cite{Donders:1869} in the very early days of experimental psychology, reaction time ({RT}) has been among the most widely used measures of cognitive processing in human and animal behavioral experiments. Very generally speaking, following Donders'  seminal work, the logic underlying the analysis of data in {RT} experiments is that, information processing takes time, thus the average time taken to initiate or complete a task reflects the duration of the process(es) that are involved in the task. Therefore, if certain types of stimuli, tasks, or groups of subjects elicit longer {RT}s than others, it is generally inferred that the former involve more cognitive processing than the latter. In this study, I propose a qualitatively different perspective on the understanding of {RT} data: Rather than focusing on whether some experimental conditions elicit shorter or longer RTs than others, I investigate whether different conditions elicit {RT} distributions with different degrees of \emph{complexity}. As I will argue, an increase in the complexity of the {RT} distribution constitutes an indirect measure of the amount of information processing that has been performed by the system.  For this, I take a psychologically naive, model-free, approach: Instead of guiding the RT analysis using knowledge about the relevant neural and/or psychological processes that give rise to RTs, I intend to draw inferences on the former by studying only the properties of the latter. 

The cognitive system can be considered a system in the thermodynamical sense of the word. In particular, it is an open system that exchanges energy (and information) with its environment. Performing an experimental task involves an exchange of information with the environment. The experimental instructions and the presentation of stimuli are a source of information. Performing the experimental task requires the processing of this external information, and information processing is costly in energy terms. As discussed by Brillouin \cite{Brillouin:1956}, the acquisition of information by any part of a system must be offset with a decrease of information somewhere else. In Brillouin's terms there is a balance between gained and lost `negentropy', that is, information. Having received energy and information, the stimulus is processed and a response is initiated. Once more, this process involves a further exchange of negentropy and energy with the environment. An ideal system with perfect efficiency could perhaps achieve a perfect balance between the received, and the spent negentropy. However, as the efficiency is never perfect, some negentropy will be lost in the process. Eventually, in the case of the cognitive system, this loss of negentropy can be compensated for by a supply of energy, normally by metabolic means, that would enable the system to return to its `resting' state. In short, the processing of experimental stimuli should temporarily increase the entropy of the cognitive system by an amount directly proportional to the amount of information that has been processed, corresponding to the  negentropy that was wasted in the process. In essence, a measure of the increases in the entropy of the cognitive system elicited by different experimental conditions or stimuli would provide an estimate of the amount of information\footnote{In this study, I follow \cite{Brillouin:1956}'s interpretation equating negentropy and information.} that has been processed (see \cite{Kirkaldy:1965} for a detailed physical description of this type of processes).

Measuring the overall state of entropy of the cognitive system might not be an easy task, as it would involve a quantification of the uncertainty in the state of all the microscopic units in the system. However, collateral measures of the `noise' emitted by the system should reflect increases in its state of complexity. This is to say, if the system is in a higher state of complexity, the noises it emits will also increase in their complexity. The random variability of times at which responses happen in a particular experimental condition can be considered as part of this emitted `noise'. Therefore the uncertainty ({\em i.e.,} entropy, in its statistical sense \cite{Shannon:1948}) of this distribution can be taken to reflect the state of the system that generated them. My working assumption is that one can measure the entropy of the distribution of RTs in a particular condition (i.e., the \emph{temporal entropy}), and make inferences about variations in the entropy  (in its physical sense) of the underlying system. In short, an increase in the informational entropy of an RT distribution is directly proportional to the amount of information that has been processed. 

In a typical repeated measures RT experiment, the differential entropy \cite{Shannon:1948} of the RT distribution can be expressed as a mixed effect model ({MEM}; see \cite{Bates:2005,Baayen:2007,Baayen:Davidson:Bates:2008} for recent introductions to this technique) with meaningful (and thus very constrained) parameter values:\footnote{See supplemental materials for the derivation of this equation.}
\begin{align}
\mathrm{E}\left[-\log p\left(t\right)\right] & = h_0 + k \sum_{i=1}^{N}\big[ \theta_i I_i(S,P) \big] + \varepsilon \nonumber \\ 
 & \qquad +  \mathrm{E}_S\left[-\log p\left(t\right)\right] + \mathrm{E}_P\left[-\log p\left(t\right)\right].
\label{eq:regression-main}
\end{align}
In this model, the independent variable is the self-information of the RTs (i.e., $-\log p(t)$), whose expected value is -- by definition -- the entropy. The intercept of the model ($h_0$) corresponds to the baseline entropy of the RT distribution, which must always be positive and provides an indication of task complexity. The fixed effect coefficients ($k \theta_i$) indicate the relative contribution of the $i$-th known source of information in the stimuli ($I_i$), and must all be positive and smaller than or equal to one. In this product, the $\theta_i$ represent the proportion of the $i$-th source of information that is processed. On the other hand, $k$ is constant for all sources of information representing the proportion of the wasted negentropy that is reflected in the RT variability. Therefore both $k$ and the $\theta_i$ must also lie within the $(0,1]$ interval. This has the additional implication that $k$ is bound to be larger than or equal to the largest observed fixed effect value, least the estimated value for some of the $\theta_i$ would be greater than one. The last two terms on the right-hand side of Eqn.~\ref{eq:regression-main} correspond to random effects of the individual stimulus $S$ and participant $P$. These correspond to other unknown sources of information linked to the identity of the stimulus or participant that are not accounted for by the $I_i$. Finally, $\varepsilon$ accounts for the error in the estimations. If estimates of $p(t)$ and of $I_i(S,P)$ can somehow be obtained, this relationship can be tested directly.

Information Theory has a long history in the study of behavior, particularly so in the study of RTs. Very soon after Shannon's development of information theory in telecommunications \cite{Shannon:1948}, psychologists were applying it to the study of human RTs. This produced one of the few standing laws of experimental psychology: The time it takes to make a choice is linearly related to the entropy of the possible alternatives; this is now referred to as Hick's Law  \cite{Hick:1952,Hyman:1953,Hellyer:1963}. Despite some possible corrections
(see, e.g., \cite{Longstreth:ElZahhar:Alcorn:1985,Welford:1987,Longstreth:Alcorn:1987}), the main claim of this law is still accepted today. More recently, a direct relation between the perceptual information content of stimuli and RTs has been found in psychophysical studies \cite{Norwich:Seburn:Axelrad:1989,Norwich:1993}. Interestingly, it is suggested that the rate at which information is collected could vary with the intensity of the stimulus \cite{Norwich:1993}. Indeed, for higher cognitive functions -- language in particular -- evidence has been presented recently that the rate of information processing is not constant, rather it is linearly related to the average informational load of the stimuli (i.e., words) in a particular experimental context \cite{Kostic:2005}. This seems to go against the spirit of Hick's Law in that information processing speed might not be constant within an experiment after all. In this study the focus is shifted from the amount of information that is \emph{carried} by the stimuli, to the amount of information about them that is actually \emph{processed}. As I will show, this change of perspective has important implications for the theoretical and empirical validity of Hick's Law.

Here, I investigate the usefulness of the thermodynamical argument to understand behavioral data. I focus on lexical stimuli, as these are less amenable to informational content measurements than plainly perceptual ones (but see also \cite{Kostic:2005,Kostic:1991,Kostic:Markovic:Baucal:2003,McDonald:Shillcock:2001,Moscoso:Kostic:Baayen:2004,Moscoso:2007,Filipovic:2007,Milin:Filipovic:Moscoso:2009} for approaches to quantifying different aspects of lexical complexity using information-theoretical measures). The empirical confirmation of the theory makes use of two large sets from the English Lexicon Project database (ELP; \cite{Balota:etal:2007}) of English visual lexical decision ({VLD}) and word naming ({WN}) data. The empirical evidence consists of four analyses. The first one investigates the relationship between the temporal entropy variable (i.e., complexity of RTs described above) with the traditional average RT (magnitude of RTs) interpretation of response latency data. This serves as a first validation of the plausibility of the approach, and it reveals its relation to Hick's Law. The second part of the analysis tests the power of the method to distinguish between the VLD and WN tasks, despite the great similarity of their average RTs. The third part provides a direct test of the theoretical development expressed in Eqn.~\ref{eq:regression-main} for particular properties of the stimuli. Finally, the fourth part goes to further depth about the implications of the relationship between mean RT and temporal entropy, and how these implications provide strong evidence (both theoretical and empirical) against Hick's Law.

\section*{Results}

\subsection*{Analysis I: Relation of temporal entropy to {RT} averages}

The  first question that arises is how do temporal entropies relate to the traditional reading of the RTs being short or long. For individual words, I investigated the relationship between both measures. For this, I used MEM regressions both on the plain (and logged) RTs, and on the RT self-informations. The relation between the random effect adjustments for individual words between models (corrected by the corresponding general intercept) summarized the relationship between the temporal entropy and RT. 

Fig.~\ref{fig:RT.SelfInf} plots the relation between the individual response self-informations and the corresponding RTs. Note that this relation is non-trivial; for the bulk of the responses, it follows a non-linear U-shaped pattern. Therefore one cannot directly assume a simple relation between the results of both measures. In contrast, Fig.~\ref{fig:nonlinear} compares the estimated mean RT with the temporal entropies for each word in each of the tasks. The dashed lines plot the best fitting linear regression between both measures. The figure suggests that the temporal entropies are linearly related to the average RTs.\footnote{The relationship between temporal entropy and log RT was also tested; it revealed a strongly non-linear pattern with lower explained variances.} This (apparently) linear relation is very strong, with explained variance values of 64\% in WN and up to 81\% in VLD. At first sight, this finding could be understood as a stronger restatement of Hick's Law \cite{Hick:1952,Hyman:1953,Hellyer:1963}: These studies provided evidence that the average RT to a stimulus is directly proportional to the informational content of the stimulus. The current results seem to suggest that the amount of information processing caused by the stimuli is also directly proportional to the average RT. This would indicate that there is a constant -- possibly task-dependent -- information processing speed of the cognitive system. Further support for this interpretation could come from the very similar slopes of the linear regression lines in the figure across visual lexical decision ($4.30 \pm \mathrm{SE}\, .05$ bits/s)\footnote{Note that these estimates have been rescaled into bits/s.} and word naming ($4.13 \pm \mathrm{SE}\, .07$ bits/s) indicating that this processing speed might be constant even across tasks. This would go one step further than Hick's Law by saying that the average RT would  be directly proportional to the amount of information that has been \emph{processed} (rather than just contained by the stimulus). Although this would be a suggestive interpretation, some caution needs to be taken before accepting this conclussion.  Analysis~IV will provide a more in-depth study of the issue of information processing speed and its implications for RT distributions. I now turn to an investigation of the effect of informational contents of stimulus and task complexity on the amount of information processing.

\subsection*{Analysis II: Estimating cognitive processing across two tasks}

For this analysis, I selected subsets of the {ELP} database that contained responses by multiple subjects to the same word stimuli across two tasks. This enabled direct pairwise comparison between RTs across the tasks. Responses in {VLD} have generally slightly longer latencies than in {WN}. This is corroborated in this dataset. Comparing the estimated average RTs to individual words in VLD and WN (as predicted by the MEM models above), revealed that VLD RTs are longer than in WN (average difference $19.82$ ms.; 95\% CI $[17.24,\, 22.39]$ ms.; paired $t[1985] = 15.09$, $p < .0001$ two-tailed) and so are the corresponding log RTs (average difference $.012$; 95\% CI $[.008,\, .015]$; paired $t[1985] = 6.46$, $p < .0001$ two-tailed). This is also the case for the temporal entropies, that is, the temporal entropy of a word in VLD is on average larger than the temporal entropy of the same word in WN (average difference $.75$ bits.; 95\% CI $[.73,\, .76]$ bits; paired $t[1985] = 113.42$, $p < .0001$ two-tailed).\footnote{These results were also confirmed by non-parametric Wilcoxon signed rank tests.}

The magnitude of the $t$ statistic was much larger for the difference in temporal entropies than for the RTs, whether logged or untransformed. Furthermore, the average difference between conditions was also much larger for the temporal entropies than for the RT measures: Whereas the RTs to words in VLD were 3.1\% longer with respect to their WN latencies (and down to .2\% in logarithmic scale), they were 8.5\% more complex. This suggested that temporal entropy provides a clearer differentiation between the tasks than did the magnitude of the RTs. This is depicted in Fig.~\ref{fig:cross-task}. The panels compare the estimated mean RTs (left panel), mean log RTs (middle panel), and temporal entropies (right panel) for WN (horizontal axes) and VLD (vertical axes). The grey dots correspond to individual words, and the dashed line is the identity condition. Whereas in the RT measures the difference between tasks (i.e., asymmetry with respect to the identity line) is barely noticeable, the temporal entropy measure sharply separates the two tasks. Only in 18 out of the 1,986 (less than 1\%) studied words did the temporal entropy have a higher value in WN than it did in VLD.\footnote{This difference was not due to differences in the size or number of participants between both datasets. See Supplemental Materials for details.}

\subsection*{Analysis III: Estimating the information processed about letters and words} 

Table~\ref{tab:effects} shows that the MEM models on the temporal self-informations, in both datasets, revealed main effects of the lexical and letter information variables on the temporal entropies that were fully consistent with the theoretical predictions. As predicted above, the fixed-effect regression coefficients ($\beta_i=k\theta_i$) were in all cases positive and smaller than one. Also, as indicated by the $\chi^2$ log-likelihood tests, the suggested random slopes (see Materials and Methods) constituted a significant improvement on the basic models. For comparison, the table also includes the results of running MEMs on the log RTs using the same predictors that were used with the temporal self-informations, revealing a nearly identical pattern.

The values of the main effect coefficients provided a lower-bound estimate for the value of the dimensionless coefficients $k$; recall that this task-specific constant measures the scaling between the entropy of the RTs and the entropy of the system. In the case of VLD, this lower-bound was $k \geq \max\{.0902,\,.0161\} = .0902 =  k^\star $, and for WN it was $k \geq \max\{.0295,\,.0921\} = .0921 =  k^\star $. Notice that the $k^\star$ estimates for both tasks were rather similar; in fact, if the standard errors of the estimates were also taken into account, there was no reason to believe that the estimates were at all different.

As discussed above, the $k^\star$ lower-bounds, combined with the fixed effect estimates of the regression ($\beta_i$) can be used to estimate the upper-bound for the possible contribution of one bit of information contained by the properties of the stimulus, into the amount of information about it that is actually \emph{processed} by the system ($\theta_i$). In VLD, one bit of lexical information corresponded to a mean of at most one bit of cognitive processing, while one bit of letter information resulted in a maximum of .18 bits of cognitive processing. Similarly, in word naming, one bit of lexical information elicited at most .32 bits of cognitive processing, while each bit of letter information could maximally correspond to a full bit of processing. These estimations put numbers into the intuition that lexical information is more relevant in VLD than it is in WN, and that letter identity information is more important for WN than it is for VLD. In both cases, much of the information that the stimulus contains is not at all processed, presumably because it is not useful for the task at hand.

\subsection*{Analysis IV: Is information processing speed constant within a task?}

I now return to the linear relationship between mean RTs and temporal entropy that was suggested to be taken with caution in Analysis~I. The question arises as to the implications that the arguably linear relationship in Fig.~\ref{fig:nonlinear} would have for the shape RT distribution. This question was addressed using the Principle of Maximum Entropy \cite{Jaynes:1957,Jaynes:1957b}. A Maximum Entropy analysis revealed that, under the assumption of an existing mean $\mu$ that is linearly related to the entropy, the most likely relationship between the entropy and the mean is described by:\footnote{See supplemental materials for the Maximum Entropy derivation of this equation.}
\begin{equation}
\kappa_2 \mu + \mathrm{E}[\log t] -\log \kappa_1 = a + b\mu,
\label{eq:impossible}
\end{equation}
where $a$, $b$ are the intercept and slope of the assumed linear relation, and $\kappa_1$, $\kappa_2$ are constants for a given dataset. Notice that a new term $\mathrm{E}[\log t]$ corresponding to the mean of the log RTs appeared in the relationship, without it having been assumed \emph{a priori}. This indicates that, if one assumed that there is a linear relationship between the mean RT and the temporal entropy, one should also assume that there is a linear relationship between the mean log RT and the temporal entropy. This entails a sort of probabilistic \emph{reductio ad absurdum}: It says that the temporal entropy is linearly related to \emph{both} the mean RT and the mean log RTs. That is, unless the mean RT and the mean log RT were independent of each other -- and they were not -- the relationship between mean RT and temporal entropy was most likely to be nonlinear after all, despite the seemingly linear appearance of the plots of Fig.~\ref{fig:nonlinear}.

To investigate this prediction, I performed a linear regression on the temporal entropies of individual words in each task, with both mean RT and mean log RT as co-variates (the effect of mean log RT was considered only after partialling out the effect of mean RT). As was predicted by the Maximum Entropy Analysis, both the visual lexical decision and the word naming datasets revealed significant linear contributions of mean RT (VLD: $\beta = 0.0084 \pm \mathrm{SE}\, .0002; F[1,1983]=11,444.43; p<.0001$; WN: $\beta = 0.0070 \pm \mathrm{SE}\, .0002; F[1,1983]=4,221.6; p<.0001$) together with a significant correction introduced by the mean log RT (VLD: $\beta = -2.96 \pm \mathrm{SE}\, .12; F[1,1983]=658.61; p<.0001$; WN: $\beta = -2.30 \pm \mathrm{SE}\, .12; F[1,1983]=379.7; p<.0001$). In both cases, the corrected estimate of explained variance increased by about 5\% when taking the mean log RTs also into account (from 81\% to 86\% in VLD, and from 64\% to 70\% in WN).\footnote{As shown in the supplemental materials, the Maximum Entropy method could also reveal an influence of the second moment of the distribution, if one assumed its existence. This would lead to a more realistic Weibull-type RT distribution, whose entropy would also have a linear term for the second moment, in addition to the terms in Eqn.~\ref{eq:impossible}. Additional regressions also confirmed this relation with the second moment, which added approximately a further 5\% explained variance to each dataset. I focus the discussion only on the additional contribution of the mean log RTs, as this is sufficient to introduce the necessary non-linearity while being considerably simpler (the estimation of the second moment required additional methods not discussed here). In any case the consideration of the second moments did not produce any significant change on the results to follow.}

The non-linearity is summarized by the additional solid lines in Fig.~\ref{fig:nonlinear}. These estimate how the correction introduced by considering the mean log RT looked like.\footnote{The lines were obtained as non-parametric smoothers between the estimated mean RTs and the predicted values of the multiple regressions using both mean RTs and mean log RTs as co-variates.} Both datasets showed a pattern reminiscent of a hockey stick, in which there were small bendings at the bottom of the ranges that became increasingly straight with increasing mean time. This was due to the decreasing importance of the log RT relative to the untransformed RT that grows much faster. 

The adaptive average processing speed is further illustrated in Fig.~\ref{fig:gradients}. The figure plots, for each of the tasks, the evolution of the average instantaneous information processing rate, as a function of the total amount of processing. The average instantaneous rates were obtained by taking the numerically estimated gradients of the solid lines in Fig.~\ref{fig:nonlinear}:
\begin{equation}
\bar{r}_t = \frac{\partial h}{\partial \mathrm{E}[t]}.
\end{equation} 
The figure shows that, in both tasks, the information processing rates increased monotonically with amount of processing, this is, the more the amount of processing that was needed, the faster it happened. In addition, there seemed to be three linear regimes for this increase that were rather similar across tasks. It is possible that the third regime, corresponding to the slower slope lines in the high values of temporal entropy were at least partly a consequence of the truncation point at 4,000~ms. Slow words will be more likely to be affected by this truncation, as a proportionally larger part of their density mass was chopped off, leading to possible underestimations of both the mean RT and the temporal entropy.

Finally, the lower-bound estimates $k^\star$ in each task obtained in Analysis~III, open the possibility of guessing the upper-bounds for the range of variation of the overall cognitive information processing speed in each of the tasks, as a function of the rate in terms of temporal entropy:
\begin{equation}
\bar{r} =  \frac{\bar{r}_t}{k} \leq  \frac{\bar{r}_t}{k^\star} = \frac{1}{k^\star}\, \frac{\partial h}{\partial \mathrm{E}[t]} = \bar{r}^\star.
\end{equation} 
In VLD this gave a range of upper-bound values ($\bar{r}^\star$) going from around 33 bits/s to about 61 bits/s, and in WN the range went between 22 bits/s and 60 bits/s. 

\section*{Discussion}

This study has introduced a novel interpretation for RT experiments. The conventional approach is to consider how long responses take to occur. Instead, I proposed to investigate whether the temporal distribution of responses is more or less \emph{complex}. I argued that this complexity of the reaction time distribution reflects the underlying state of complexity in the cognitive system, and the empirical evidence has supported this view. This enables a shift from studying how much information is contained in stimuli or tasks, to directly investigating the amount of this information that is actually processed.

As evidenced by the comparison between VLD and WN, the temporal entropy measure is remarkably more sensitive than the traditional RT magnitudes -- whether in untransformed or logarithmic scale -- in distinguishing tasks with different properties. There has been a growing interest in techniques that enable going beyond the mean in the description of RT data \cite{Ratcliff:1978,Luce:1986,Heathcote:Popiel:Mewhort:1991,VanZandt:2002,Rouder:etal:2005,Balota:etal:2008,Holden:VanOrden:Turvey:2009}. These proposals consist in studying either the quantiles or the higher moments of the distribution, or the parameters of some distribution family that is assumed \emph{a priori}, which are estimated separately for different participants and/or experimental conditions. I have proposed a considerably more simple, and model-free, measure: The entropy of the RT distribution summarizes the cognitively  relevant aspects of its shape. The working assumption, from an information processing perspective, is that any variation in the amount of processing must be reflected in the entropy of the distribution. By implication, the temporal entropy should be a sufficient statistic to reflect the effects of different cognitive manipulations. Furthermore, the new measure uses the random effect structure of the experiment, and is thus less sensitive to the sometimes very reduced number of points of each individual condition (see, e.g, \cite{Balota:etal:2008}).

Entropy is, by definition, an additive measure. Different contributions of independent factors can then be considered in an plainly additive manner. Current practice in the analysis of RT data recommends transforming the RTs prior to statistical analyses, either using a logarithmic or a reciprocal transform. This has the undesirable effect of breaking the additive interpretability of effects, forcing researchers to delve into complex multiplicative processes \cite{Holden:VanOrden:Turvey:2009}. In contrast, the approach proposed here remains in the additive domain, while at the same time being able to capture complex aspects of the distributional shape. This is achieved while keeping with a model-free approach, that is, no particular distributional shape needs to be assumed.

An important consequence of this analysis is the conclusion that Hick's law needs to be extended and reformulated. Strictly speaking, the law proposed by Hick, Hyman, and others \cite{Hick:1952,Hyman:1953,Hellyer:1963} concerns only the relation between information contained in the stimuli and mean RT. Crucially, in this study, I extend this argument to the information about the stimuli that is actually processed, this is to say, the relevant information. In this case, the relationship between processed information and mean RT is not linear, even though it might seem linear to the naked eye. The non-linearity was suggested by Maximum Entropy theoretical analysis and confirmed by the empirical data. Note that collectively these results strengthen the argument considerably: The non-linearity is not just obtained from a fit to the data, but was predicted \emph{a priori} in detail. Although the adjustment might seem small on the average entropies, inspection of Fig.~\ref{fig:gradients} makes it clear that that the non-linear relation allows the information processing rate to double in some contexts. This finding suggests an adaptive system where the processing load is dynamically adjusted to the task demands. In a way, it is a `lazy' system in the sense of Zipf \cite{Zipf:1949}.

Not surprisingly, the increase of information processing rate with stimulus complexity is consistent with the findings of Kostic \cite{Kostic:2005} for VLD, even in the estimates of the information processing rate. In a {VLD} task using Serbian words, Kostic estimated that, depending on the average informational content of the particular stimuli in an experiment, the information processing rate ranged from about 30 to about 100 bits/s. This is consistent with the maximum rate (i.e.,~$\bar{r}^\star$) ranging from 30 to 60 bits/s that I have derived. The difference between both estimates are possibly due to the difference between information contained in the stimuli that Kostic studied, and information that is processed about them that has been studied here. Notice also that Kostic's estimates refer to the aggregated averages over full experiments. In contrast, the technique proposed here enables obtaining estimates for individual stimuli.

The principal contribution of the present study is that the entropy of an RT distribution provides an index of the amount of information processing that has taken place. Stated alternatively, changes in the entropy of RT distributions reflect changes in the underlying state of the cognitive system. I have estimated that, in the tasks under study, a minimum of around 10\% of the increase in the entropy of the cognitive system is reflected in an increase of temporal entropy (i.e., the $k^\star$ estimates). This finding provides a handle by which RT data can be used to establish the link between higher cognitive function and its metabolic counterparts, that was proposed in the seminal study of Kirkaldy \cite{Kirkaldy:1965}. Furthermore, the characterization of behavioral responses in terms of entropy enables the consistent treatment of behavioral and neurophysiological signals using the same theoretical tools (see \cite{Borst:Theunissen:1999} for a review on Information Theory in the study of neurophysiological signals).

\begin{figure}
\begin{center}
\includegraphics[width=.8\textwidth]{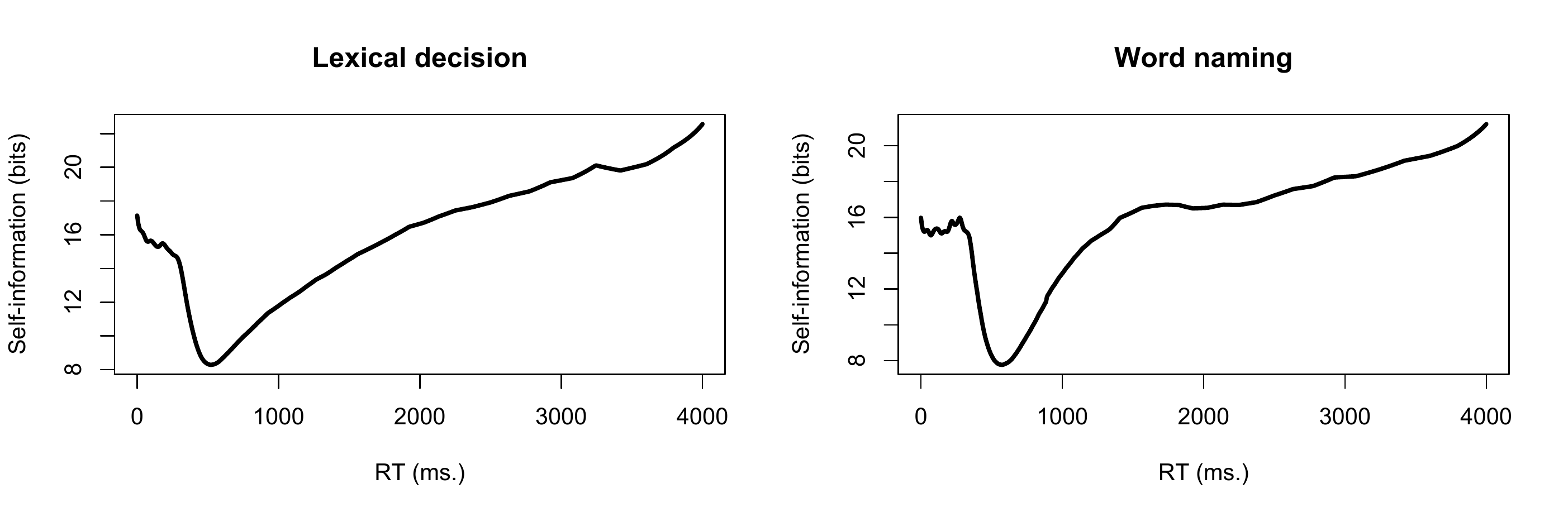}
\end{center}
\caption{Relation between reaction time and self-information. The left panel plots the Visual Lexical Decision dataset and the right panel plots the Word Naming one. Reaction time self-informations were estimated combining Gaussian {KDE} estimated at untransformed scale for the left-tails and logarithmic scale for the right tails.}
\label{fig:RT.SelfInf}
\end{figure}

\begin{figure}
\begin{center}
\includegraphics[width=.8\textwidth]{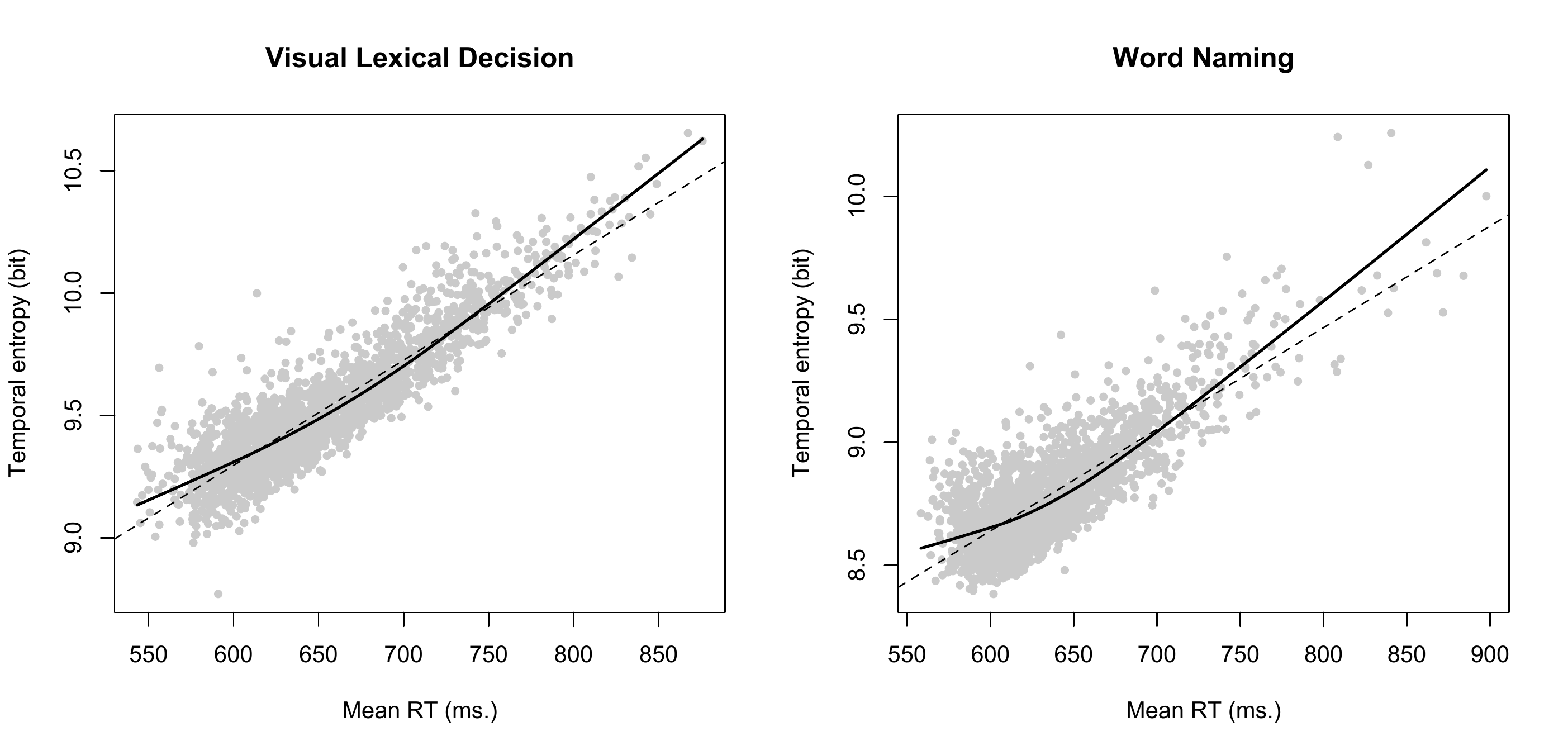}
\end{center}
\caption{Relation between average RT and temporal entropy. The left panel plots the visual lexical decision dataset, and the right panel plots the word naming one. The grey dots plot the estimates of mean RT and temporal entropy for individual words. The dashed lines illustrate the best fit of a purely linear relationship between both measures, as would be characteristic of Hick's Law. The solid lines plot this relationship when the effect of the mean log RT is also considered.}
\label{fig:nonlinear}
\end{figure}

\begin{figure}
\begin{center}
\includegraphics[width=.8\textwidth]{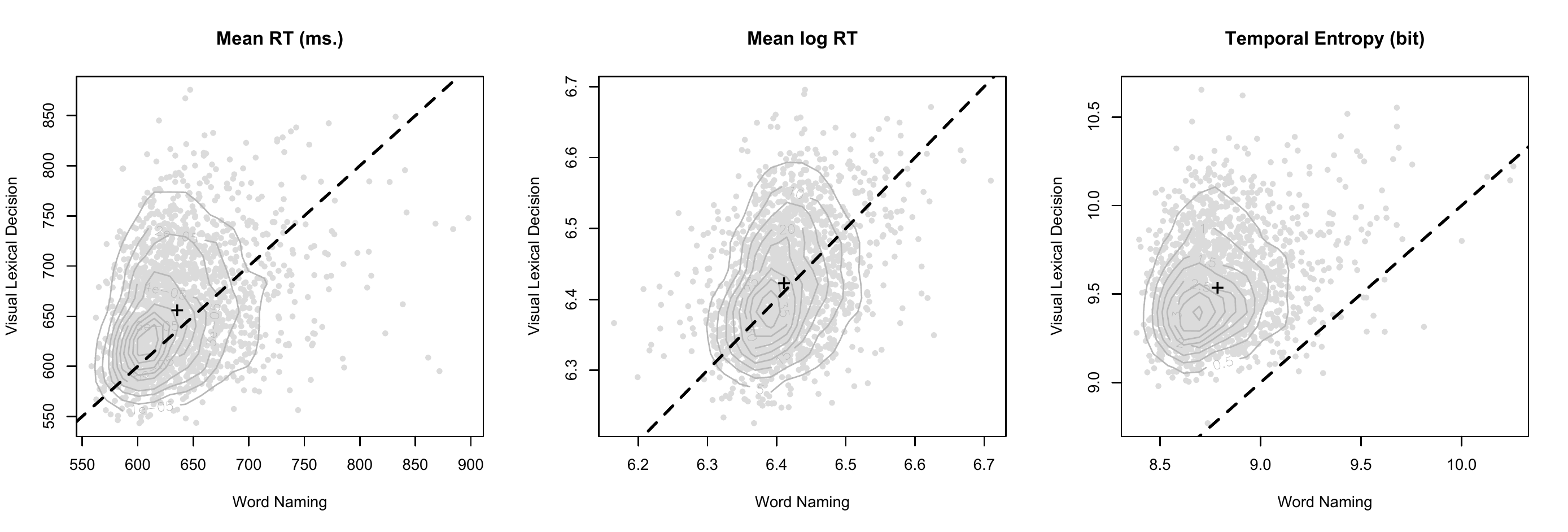}
\end{center}
\caption{Cross-task comparison.  The plots compare the estimated mean RTs (left panel), mean log RTs (middle panel), and temporal entropies (right panel) for word naming (horizontal axes) and lexical decision (vertical axes). The light grey dots correspond to individual words, and the dashed line is the identity condition. The dark grey contours corresponds to a two-dimensional KDE. The crosses plot the mean of both tasks for each of the three measures.}
\label{fig:cross-task}
\end{figure}

\begin{figure}
\begin{center}
\includegraphics[width=.8\textwidth]{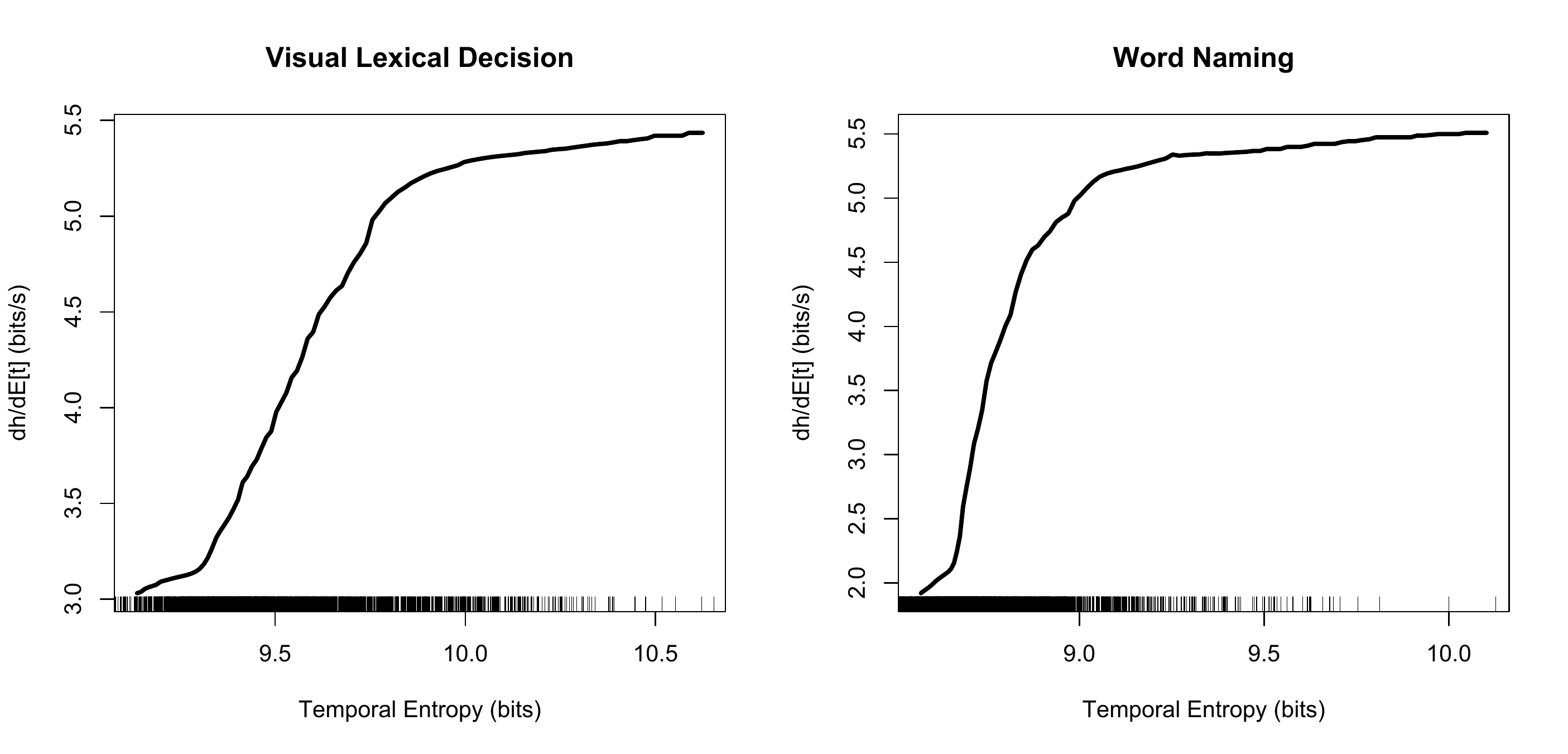}
\end{center}
\caption{Relation between information processing rate  and temporal entropy. The left panel plots the lexical decision estimates, and the right panel plots the word naming one. Both panels plot temporal entropy on their horizontal axes, and an estimate of the average instantaneous processing rate ($\bar{r}_t$) in the vertical axis. The estimates were obtained as the numerical gradients of the solid black lines in Fig.~\ref{fig:nonlinear}. The rugs at the bottom of the panels illustrate the approximate number of points on which each portion of the curve is estimated.}
\label{fig:gradients}
\end{figure}

{\footnotesize
\section*{Materials \& Methods}

\subsection*{Materials}

I retrieved from the {ELP} database \cite{Balota:etal:2007} the individual lexical decision {RT}s and word naming latencies to 1,986 nouns and verbs. The set of selected words corresponds to those words used in a previous study that compared the lexical decision and naming tasks \cite{Baayen:Feldman:Schreuder:2006}. The selection of only a subset of the words enables us to keep the models below tractable. All responses that were not marked as correct in the database were excluded from further analysis. In addition, I also excluded all responses equal or longer than 4,000 ms., as beyond this limit responses appear truncated in the {ELP} database. This left a total of 64,087 responses (from 816 different participants) in the lexical decision dataset and 53,403 (from 445 participants) in the naming one. Along with the individual RTs, the surface frequency of the words (extracted from the {CELEX} database \cite{Baayen:CD95}) and their length in characters was also recorded. The word frequency and word length measures were transformed into information theoretical measures, the self-information of the word, and its average informational content due to its letters.\footnote{See supplemental materials for details on these transformations.}

\subsection*{Estimation of the individual {RT} self-informations}

The individual self-information values for each {RT} in the visual lexical decision and naming datasets were estimated by Kernel Density Estimation (KDE; \cite{Parzen:1962}) with Gaussian kernels, and imposing bounds on the distribution at the truncation points of 0~ms. and 4,000~ms., beyond which the density was estimated to have a value of zero (with the adequate normalization to integrate to one in that interval). The estimates combine a direct {KDE} for the left tail of the distribution, and a retransformed estimate on logarithmic scale for the right tails. The reason for this dual estimate was to avoid the high-frequency noise that {KDE} introduces on the right tails of heavy-tailed distributions. Estimating the distribution in logarithmic scale greatly attenuates the noise on the right tail \cite{Newman:2005}, but it introduces additional noise in the vicinity of zero, thus the dual estimation. To guarantee a certain smoothness in the transition between the untransformed and the logarithmic KDE, in the area around the distributional mode I used a weighted average between both estimates. From the total support grid of 512 points where the densities were estimated, 40 points  to the left of the mode received an weighting linearly decreasing in favor of the transformed estimate. This pattern was reversed in the 40 points to the right of the mode, where the weighting was linearly increased to favor the logarithmic estimates. The mode itself received a plain .5/.5 average between both estimates.

The individual self-information values correspond to the the minus logarithm of the estimated density, using two as the base of the logarithm to obtain an estimate in bits. The individual values for each response were interpolated from the 512 point grid on which the densities were estimated. Fig.~\ref{fig:RT.SelfInf} shows the result of this process for both datasets under study.

\subsection*{MEM regressions}

To investigate the relationship between temporal entropy and mean RTs (Analyses~I, II, \& IV), I fitted two MEM models to the data. In both cases, the models included an intercept, and two random effects, one of word and one of subject, as predictors. The first model had the RT as dependent variable (so that the average predictions of the model correspond to the mean RTs), and the second model had the self-information of the RTs as independent variables (thus as argued above, the average predictions of the model correspond to temporal entropies). The individual estimates of either mean RT or temporal entropy for each word were computed as the sum of the corresponding model's intercept with the particular random effect adjustment for that word.

To investigate the predictions on the informational content of stimuli (Analysis~III), I performed MEM regressions to the temporal self-information and to the log RTs in each task. As before, these regressions included random effects of both subject and word. In addition, I also included fixed effect covariates measuring the lexical self-information and the average information content of its letters. An additional factor needed to be considered in these MEMs. The measure of the informational content of the letters in a word refers to the average case. However, different words will contain combinations of letters that contain more or less information than the average (i.e., they are either more frequent than usually or rarer than usual). Therefore, the effect of the information of the letters can vary with respect to word identity. To address this problem, the MEM models included the possibility of a random slope making the effect of letter-based information variable for different words. Similarly, the experience that different people have had with different words varies in both quantitative and qualitative terms. This was accounted for by introducing a random slope that enables the variation of the lexical self-information effect across individual participants.

All {MEM} regressions were fitted using a restricted maximum likelihood algorithm, as implemented in the \emph{R} package ``lme4'' \cite{Bates:2005}.

\section*{Acknowledgments}
The author is indebted to L.~B. Feldman, D. Filipovi\'c-{\DJ}ur{\dj}evi\'c, I.~J. Grainger, and M. Mondrag\'on for helpful suggestions.
}

\newpage

\begin{table}
\caption{Summary of Mixed-Effect Model Results. Effects of lexical self-information and letter information on the temporal entropy and log RTs as estimated by the MEM models. The upper section of the table provides the estimates of the fixed effects and their associated statistics. The lower section provides the model comparison statistics (log-likelihood tests) comparing models including different combinations of random slopes.
$^1$ The estimation of the degrees of freedom for an MEM is not a straightforward issue and the $p$-values can be considered as slightly lax (cf., [6]). However, for so large degrees of freedom, the possible underestimation of the $p$-values would be negligible.
$^2$ The effect of letter information in visual lexical decision is only marginally significant (in a two-tailed test). However it was significant in the basic model without random slopes  ($\beta = .0175 \pm \mathrm{SE}\, .0079;\, t[53,400] = 2.21,\, p=.0271$). The log-likelihood tests recommended keeping the associated random slope and thus also the fixed effect.
}
\resizebox{\textwidth}{!}{
\begin{tabular}{c|cc|cc}

\hline
Dependent variable:  & \multicolumn{2}{c}{Temporal Entropies} & \multicolumn{2}{c}{Log RTs} \\
\hline
					& Lexical Decision & Word Naming & Lexical Decision & Word Naming \\
\hline
\hline
\multirow{3}{*}{Lexical Self-inform.}	& $\beta = .0902 \pm \mathrm{SE}\, .0042$ & $\beta = .0295 \pm \mathrm{SE}\, .0044$ & $\beta = .0273  \pm \mathrm{SE}\, .0008$ & $\beta=.0086 \pm \mathrm{SE}\, .0008$\\
 & $t[64,084] = 21.48$\footnotemark[1] & $t[53,400] = 6.74$ & $t[64,084] = 35.40$ & $t[53,400] = 11.20$ \\
 & $p<.0001$               & $p<.0001$ & $p<.0001$ & $p<.0001$ \\
\hline
\multirow{3}{*}{Letter Information} & $\beta = .0161 \pm \mathrm{SE}\, .0084$ & $\beta = .0921 \pm \mathrm{SE}\, .0093$ & $\beta=.0093  \pm \mathrm{SE}\, .0018 $ & $\beta=.0273 \pm \mathrm{SE}\, .0018$ \\
 &  $t[64,084] = 1.91$ & $t[53,400] = 9.89$ & $t[64,084] = 5.20$ & $t[53,400] =14.90$ \\
 &  $p=.0561$\footnotemark[2] & $p<.0001$ & $p<.0001$ & $p<.0001$ \\
\hline
\hline
Random Slope: & $\chi^2[2] =11.25$ & $\chi^2[2] =79.90$ & $\chi^2[2] = 6.49$ &\multirow{2}{*}{$p=1$}\\
Lett. Info. by Word & $p=.0036$        & $p<.0001$ & $p=.0389$ &\\
\hline
Random Slope: & $\chi^2[2] =499.34$ & $\chi^2[2] =111.28$ &  $\chi^2[2] = 119.60$ &  $\chi^2[2] = 32.67$ \\
Lex. SI. by Subject & $p<.0001$        & $p<.0001$ & $p<.0001$ & $p<.0001$ \\
\hline
Random Slopes & $\chi^2[2] =9.89$ & $\chi^2[2] =65.65$ & $\chi^2[2] = 6.59$ & $\chi^2[2] = 11.59$ \\
(both simultaneously) & $p=.0071$ & $p<.0001$ & $p=.0370$ & $p=.0030$\\
\hline
\end{tabular}
}

\label{tab:effects}
\end{table}

\clearpage

\renewcommand{\thepage}{S-\arabic{page}}
\renewcommand{\thetable}{S-\arabic{table}}
\renewcommand{\theequation}{S-\arabic{equation}}
\renewcommand{\thefigure}{S-\arabic{figure}}
\setcounter{equation}{0}
\setcounter{page}{1}

\appendix 

\section*{Supplemental Materials}

\subsection*{Derivation of the MEM regression}

If $t$ are the times of responses elicited in an condition $C$ that requires an amount of information processing $I_P(C)$, it can be predicted that:
\begin{equation}
k I_P(C) \simeq h(t) - h_0 \geq 0,
\label{eq:prediction1}
\end{equation}
where $h_0 > 0$ is the differential entropy [1] of the RT distribution in the resting state, $k \in (0,1]$ is a constant indicating the proportional reflection of the increase of the system's entropy into the {RT} distribution, and  $h(t)$ is the differential entropy of the distribution of RTs:
 \begin{equation}
h(t)  =  -\int_0^\infty p(t) \log p(t) \mathrm{d}t. 
\label{eq:entropy-integral}
\end{equation}
The inequality in the right side of Eqn.~\ref{eq:prediction1} results from the fact that the in order to process the information, the system forcefully needs to have increased its entropy ($h(t) \geq h_0$).

Information processing is the result of an experimental situation. In a particular task, the amount of processing required will be different for different stimuli. Therefore, one can also consider a task-specific informational content of the stimuli themselves, which is external to the cognitive system. The amount of information processing involved in processing a particular experimental condition is bound to be lower or equal to the total information content of the stimulus, this is to say, the amount of information that is available limits the amount of information that can be processed. Denoting this external informational content of a particular stimulus in a given task by $I_S(C)$ we see that:
\begin{equation}
I_S(C) \geq I_P(C) \geq 0.
\label{eq:limit}
\end{equation}
Furthermore, if the estimate of the stimulus information content is relatively accurate, the amount of information that is processed should be proportional to what is available:
\begin{equation}
I_P(C) = \theta I_S(C),
\label{eq:proportion}
\end{equation}
where $\theta \in (0,1]$ represents the proportion of available information that is processed. Combining Eqns.~\ref{eq:prediction1}, \ref{eq:entropy-integral}, and \ref{eq:proportion} one obtains:
\begin{equation}
\theta I_S(C) \simeq \frac{1}{k}\left( -\int_0^\infty p(t) \log p(t) \mathrm{d}t - h_0 \right) .
\label{eq:prediction2}
\end{equation}

Taking the $k$ proportion to be relatively constant across participants in a given experimental context, in principle, the relationship in Eqn.~\ref{eq:prediction2} could be tested experimentally, providing a direct measurement of the information processing involved in a task across conditions. However, direct application of these expressions to actual experimental data is problematic. The value of the $h_0$ term is unknown, and it does not seem easy to estimate it (but see also [2]). In addition, the actual distribution of RTs ($p(t)$) is itself unknown, only a particular sample of RTs obtained in an experiment is available, and it is in most cases rather sparse.

The first problem is circunvented by studying the \emph{relative} increase in RT entropy elicited by several experimental conditions $C_1, C_2, \ldots, C_n$. Between any two given conditions $C_i$ and $C_j$. In this case, from Eqn.~\ref{eq:prediction2} one finds that:
\begin{equation}
\theta \left[ I_S(C_i) - I_S(C_j) \right] \simeq \frac{1}{k} \left[ h\left( t_i\right) - h\left( t_j\right) \right],
\label{eq:difference}
\end{equation}
where $t_i$ and $t_j$ refer to the RTs obtained in conditions $C_i$ and $C_j$. This can be readily extended to a regression situation in which the informational content of the stimuli is varied continuously. In such a case, if $t_C$ represents the RTs for a particular stimulus $C$:
\begin{equation}
\alpha + \theta \beta I_S(C) \simeq \frac{h\left(t_C\right)}{k}.
\label{eq:regression}
\end{equation}
Note that the $\alpha$ and $\beta$ coefficients above are themselves meaningful. On the one hand, the intercept coefficient $\alpha$ reflects the baseline level of temporal entropy scaled up by the proportionality  constant ($h_0/k$).This implies that one can force $\alpha > 0$. On the other hand, the slope coefficient $\beta$ corresponds to the increase in information processing per processed unit of information. This is so because the $\theta$ and $k$ coefficients already index what proportion of the stimulus information is processed, and how this processing relates to the RT distribution entropy. Therefore, trivially, $\beta = 1$. Including an explicit error term $\varepsilon$ results in:
\begin{equation}
h_0 + k \theta I_S(C) + \varepsilon = h\left(t_C\right) ,
\label{eq:regression-final}
\end{equation}
with $h_0 > 0$ and $k, \theta \in (0,1]$. This provides a direct route to testing the relationship. A linear regression with the informational content of the stimuli as predictor and the entropy of the corresponding RT distribution as dependent variable provides a direct estimate of the parameters (scaled by $k$). It is worth noting here that Eqn.~\ref{eq:regression-final} would also enable reasoning in the opposite direction. Given an estimate of the cognitive cost of processing different stimuli, one could also obtain an estimate of their relevant informational load.

The second problem concerns the estimation of the entropy of the RT distribution in a particular condition ($h(t_C)$). In a typical repeated measures experimental design, several participants respond to to different stimuli. In these cases the entropy of the RT distribution is determined not only by the known informational content of the stimuli, but also by differences on the entropy of the RT distributions of particular participants, and by additional informative issues of the stimuli that are unknown to the experimenter or difficult to control for. Thus the entropy on the overall RT distribution is the sum of multiple sources of uncertainty:
\begin{equation}
h(t) = h(t_C) + h(t_S) + h(t_P),
\end{equation}
where $h(t_S)$ is the RT entropy that is intrisical to the particular stimulus $S$, and $h(t_P)$ is the RT entropy of a particular participant $P$. Taking this into account, for a response of an individual participant $P$ to a stimulus $S$, we need to extend (\ref{eq:regression-final}) to:
\begin{equation}
h_0 + k \theta I_S(C) + h(t_S) + h(t_P) + \varepsilon = h(t).
\label{eq:regression-mixed}
\end{equation}
By definition, the entropies can be reformulated in as the expected values of the negated log-probabilities (i.e., the self-informations; [1]) of the RTs. Considering simultaneously the effects of $N$ independent known sources of information: 
\begin{equation}
h_0 + k \sum_{i=1}^{N}\left[\theta_i I_S(C)\right] +  \mathrm{E}_S\left[-\log p\left(t\right)\right] + \mathrm{E}_P\left[-\log p\left(t\right)\right] + \varepsilon =  \mathrm{E}\left[-\log p\left(t\right)\right].
\label{eq:regression-mixed2}
\end{equation}
This corresponds to the expression of a regression model with an intercept, a covariate $I_S(C)$, and two random effects $S$ and $P$, with the self-information of the individual RTs as dependent variable. The parameters of the regression have a direct interpretation. The intercept corresponds to the baseline entropy of the system for the task at hand, it is thus a measure of overall task complexity. The fixed effect coefficients correspond to the $k \theta_i$ products, that is, the influence of the known informational content of the stimulus on the amount of processing, weighted by the proportion of processing that is reflected in the increase in RT complexity. In general, to ensure that $\theta_i \in (0,1]$ it must hold that $k \geq \max \{ \hat{\beta_i} \} = k^{\star}$, where $\hat{\beta_i}$ are the estimated fixed effect coefficients of the regression. This provides a useful lower-bound for this parameter. Furthermore, the limitations above also force that all fixed effect coefficients must fall in the range $(0,1]$.

\subsection*{Informational Content of the Stimuli}

The word length counts were transformed into an informational measure using a corpus based estimate of the average entropy rate of English of 1.23 bits per letter [3] discounting an estimate of .04 bits per characters estimated to reflect the information carried by spaces or case information [4].\footnote{[4] actually estimated .06 bits per character for spaces and case, but this was scaled down to account for the difference in the overall estimate with the [3] estimates, which are considered the best available approximations.} Thus, the information content of a word due to its letters was estimated as.
\begin{equation}
I_L(w) = 1.19 \cdot l(w),
\label{eq:inforlen}
\end{equation}
where $l(w)$ is the word length in letters of the word $w$.

Different words in a language vary with respect to the amount of information that they convey. Generally speaking, to quantify the precise amount of information that is conveyed by a word seems at best very difficult. However, coming up with a theoretical, \emph{a-priori} estimate of the information contained by a word would require the joint consideration of multiple linguistic and contextual factors, many of which are yet poorly understood. Here, I only consider one simple measure of a word's informativity, its self-information derived from its frequency:
\begin{equation}
I_F(w) = \log \frac{1}{f(w)} = -\log f(w),
\label{eq:selfinfo}
\end{equation}
where $f(w)$ is the relative frequency of occurrence of the word $w$ in the {CELEX} database [5].

\subsection*{Maximum Entropy Analysis}

What does the knowledge that there is a linear relationship between the mean RT and the temporal entropy tell us about the distributional shape? The less biased or more reasonable distributional shape to believe in is the one that satisfies the constraint while introducing as little additional knowledge as possible [6,7]. If $m(t)$ is the distribution that reflects our full ignorance about the possible values of the {RT}, one should choose a new distribution $p(t)$ that satisfies the constraints while being as similar as possible to the `know-nothing' distribution $m(t)$. Mathematically, this is given by the Shannon-Jaynes entropy, that is, the Kullback-Leibler divergence [8] between $p(t)$ and $m(t)$:
\begin{equation}
h_{S-J}\left( p \Vert m \right) = -\int_{0}^{T_{\max}} p(t) \log \frac{p(t)}{m(t)} \,\mathrm{d}t. \label{eq:shannon-jaynes}
\end{equation}

In his introduction of the Transformation Groups argument, Jaynes derived the shape of the full ignorance prior for the rate parameter $r$ of a Poisson distribution: \begin{equation}
m_r(r) \propto \frac{1}{r}. \label{eq:Jaynes-Poisson}
\end{equation}
The justification for the necessity of this choice comes from a general consistency argument. Consider two separate observers who were to assign probabilities to the rate of occurrence of an event. The two observers use different mechanisms to measure time, using perhaps different units (e.g., one uses milliseconds and the other uses minutes). If both observers are fully ignorant of the nature of the process, the most reasonable thing would be that they would asign an \emph{a priori} probability distribution that reflects their complete ignorance. By ignorance it is meant that the observers know strictly \emph{nothing} about the process that generates these events further than that they might happen with a rate of occurrence equal or greater than zero. Obviously, as the level of ignorance of the observers is equivalent, any consistent prior distribution would be one by which both observers assign exactly the same probability distribution to the rate, irrespective of the measuring units they each use. This requires a prior probability that is in accord with Eqn.~\ref{eq:Jaynes-Poisson}. Note that the prior distribution in Eqn.~\ref{eq:Jaynes-Poisson} is an improper one: It cannot integrate to one in the domain $[0,\infty)$. However, in Bayesian and Maximum-Entropy analyses this does not constitute a problem, as only the posterior needs to be normalized (cf., [9]). Furthermore, the recorded RTs in any experiment have a practical upper-bound at some time $T_{\max}$, and in such cases the proposed prior is proper.

The argument of [10] can be readily extended to obtain a full ignorance prior for the times at which events occur; one that can then be used for the analysis of RT distributions. The rate at which events occur is the reciprocal of the times at which they happen ($r = 1/t$).
Therefore, knowing the prior distribution for the rate, one can directly infer the prior for the times themselves, such that both priors are consistent with each other (e.g., in the example above, a third ignorant observer might have decided to infer the rates from the times, and his state of ignorance must be equivalent to that of the other two observers):
\begin{equation}
m_t(t) = m_r(r) \left| \frac{\mathrm{d}r}{\mathrm{d}t} \right| =\frac{1}{t^2} \, m_r\left(\frac{1}{t}\right)  = \frac{c\,t}{t^2} = \frac{c}{t}.
\end{equation}
where $c$ is the constant part of the prior distribution of the rates. In sum, the ignorance prior for the times must be the same as the ignorance prior for the rates.

The problem is then to maximize Eqn.~\ref{eq:shannon-jaynes} subject to the constraints:
\begin{align}
\int_{0}^{T_{\max}} p(t)\,\mathrm{d}t & =  1 \label{eq:constc0}\\[6pt] 
\int_{0}^{T_{\max}} p(t)\, t \,\mathrm{d}t & =  \mu \label{eq:constc1}\\[6pt]
-\int_{0}^{T_{\max}} p(t)\log p(t) \,\mathrm{d}t & =  a + b\mu \label{eq:constc2}
\end{align}
Constraint~\ref{eq:constc0} is the usual normalization requirement for proper distributions, \ref{eq:constc1} represents the assumption of an existing finite mean RT $\mu$, and \ref{eq:constc2} expresses the proposed linear relation between the mean RT and the temporal entropy.

This is an optimization problem that can be solved using the method of Lagrange multipliers from variational calculus. This results in a distribution of the form:
\begin{equation}
p(t) = \kappa_1 \frac{1}{t} \mathrm{e}^{-\kappa_2 t}, \qquad \kappa_1,\kappa_2 > 0,
\label{eq:maxentd1}
\end{equation}
which is a power law (with exponent -1) with an exponential cutoff. Plugging Eqn.~\ref{eq:maxentd1} into the linear relation constraint of Eqn.~\ref{eq:constc2}, and simplifying using Eqn.~\ref{eq:constc0} and Eqn.~\ref{eq:constc1}, one finds that:
\begin{equation}
\kappa_2 \mu + \mathrm{E}[\log t] -\log \kappa_1 = a + b\mu,
\label{eq:impossible}
\end{equation}

It is important to notice here that the distribution in Eqn.~\ref{eq:maxentd1} is in fact a rather implausible one for RT distributions; it is monotonically decreasing. The argument is not that this is the best, or even a good, distribution to account for RTs, but rather that it is the most reasonable one to believe in if one assumed \emph{only} the information that was given. Including further knowledge about the distribution in the form of additional constraints will produce a distribution that is more and more similar to the actual RT one. As an example, consider that one also assumed that the distribution has a known variance (which would also be safe assumption provided the RTs are truncated at some $T_{\max}$).

Including also information on the second moment of the distribution amounts to adding one further constraint to Eqns.~\ref{eq:constc0}, \ref{eq:constc1}, and \ref{eq:constc2}:
\begin{equation}
\int_{0}^{T_{\max}} p(t) t^2 \,\mathrm{d}t  \; =  \; \xi,\label{eq:newconst}
\end{equation}
where $\xi$ is the finite value of the second moment. In this case, the resulting distribution would be:
\begin{equation}
p(t) = \kappa_1 \frac{1}{t} \mathrm{e}^{\kappa_2 t + \kappa_3 t^2},
\label{eq:maxentd2}
\end{equation}
and the relation between mean and entropy would now be:
\begin{equation}
\kappa_2 \mu +\kappa_3 \xi + \mathrm{E}[\log t] -\log \kappa_1 = a + b\mu.
\label{eq:impossible2}
\end{equation}
Notice that the mean log RT term is still present. The origin of this term lies in the ignorance prior itself. Threfore, even if one included many additional constraints, such as futher higher moments, quantile values, or actual observed values, the term will remain there.

\subsection*{Further Analyses on Task Differentiation}

As the VLD dataset contained more responses per word that the WN one, a possibility is that the increase in entropy was due to a bias in the entropy estimates introduced by sample size. This possibility was discarded by a re-sampling analysis: Random downsampling of the VLD dataset to the same size as the WN one did not affect the results above in any significant way. Another plausible confound is that the larger number of participants in the VLD dataset (816 \emph{vs.} 445) might have led to higher heterogeneity and thus larger estimates of the temporal entropy, despite the participants having been controlled for by treating them as an explicit random effect. To investigate this possibility, I reduced the VLD dataset to include only the 445 participants that provided the highest number of responses. This resulted in a drastic reduction in the size of the VLD dataset, which went down to 38,639 responses from the original 64,087 (60.3\%), and thus became much smaller than the WN dataset (53,403 responses). Still, the results remained virtually unchanged, save for the difference in log RTs that failed to reach significance in the downsampled analysis ($p = .18$). The corresponding plots were plainly indistinguishable from those of Fig.~3 from the main text, as were the estimates of the difference between tasks. The individual word entropy estimates from the reduced dataset accounted for 76\% of the variance of the entropy estimates from the full dataset. This is specially remarkable considering that, for some of these words, the reduced dataset was left with as few as a single response.


\subsection*{Bibliography}

\begin{enumerate}
\item Shannon, C.~E. (1948) A mathematical theory of communication. {\em Bell Syst Tech J} {\bf 27}, 379--423,  623--656.
\item {Moscoso del Prado Mart\'{\i}n}, F. (2009) {\em The baseline for response latency distributions}. (Submitted manuscript). Available from {N}ature {P}recedings {\tt http://hdl.handle.net/10101/npre.2009.3622.1}
\item Rosenfeld, R. (1996) A maximum entropy approach to adaptive statistical language modelling. {\em Comp Speech \& Lang} {\bf 10}, 187--228.
\item Brown, P.~F, {della~Pietra}, V.~J, Mercer, R.~L, {della~Pietra}, S.~A,  \& Lai,
  J.~C. (1992) An estimate of an upper bound for the entropy of {E}nglish. {\em Comp Ling} {\bf 18}, 31--40.
\item Baayen, R.~H, Piepenbrock, R,  \& Gulikers, L. (1995) {\em The {CELEX} lexical database ({CD-ROM})}. (Linguistic Data Consortium, University of Pennsylvania, Philadelphia, PA).
\item Jaynes, E.~T. (1957) Information theory and statistical mechanics. {\em Phys Rev} {\bf 106}, 620--630.
\item Jaynes, E.~T. (1957) Information theory and statistical mechanics {II}. {\em Phys Rev} {\bf 108}, 171--190.
\item Kullback, S \& Leibler, R.~A. (1951) On information and sufficiency. {\em Ann Math Stat} {\bf 22}, 79--86.
\item Sivia, D.~S \& Skilling, J. (2006) {\em Data {A}nalysis: {A} {B}ayesian {T}utorial (2nd {E}dition)}. (Oxford University Press, Oxford, {UK}).
\item Jaynes, E.~T. (1968) Prior probabilities. {\em {IEEE} Trans Sys Sci \& Cyb} {\bf {SSC}--4}, 227--241.
\end{enumerate}

\end{document}